# An accuracy-enhanced transonic flow prediction method fusing deep learning and reduced-order model

Xuyi Jia, Chunlin Gong, Wen Ji, Chunna Li*

Shaanxi Aerospace Flight Vehicle Design Key Laboratory, School of Astronautics, Northwestern Polytechnical University, Xi'an 710072, China

∗ Corresponding author.

E-mail address: xuyijia@mail.nwpu.edu.cn (X. Jia)

leonwood@nwpu.edu.cn (C. Gong)

jiwen801@mail.nwpu.edu.cn (W. Ji)

chunnali@nwpu.edu.cn (C. Li)

**Abstract**

It's difficult to accurately predict the flow with shock waves over an aircraft due to the flow's strongly nonlinear characteristics. In this study, we propose an accuracy-enhanced flow prediction method that fuses deep learning and reduced-order model to achieve fast flow field prediction for various aerodynamic shapes. First, we establish the convolutional neural network-proper orthogonal decomposition (CNN-POD) model for mapping geometries to the entire flow field. Next, local flow regions containing nonlinear flow structures are identified through POD reconstruction for enhanced modeling. Then, a new CNN model is employed to map geometries to the local flow field. The proposed method is finally applied in predicting transonic flow over airfoils. The results indicate that the proposed enhanced DNN method can reduce the prediction error of flow properties, particularly in the regions with shock waves (up to 13%-46%). Additionally, the better efficiency and robustness of the proposed methods have been validated in comparison to existing methods.

***Keywords:*** Transonic flow prediction; Shock wave; Accuracy-enhanced; Reduced-order model; Enhanced deep neural network

## 1. Introduction

The design and optimization of aircrafts requires extensive computational fluid dynamics (CFD) simulations, which is time-consuming. In order to improve design and optimization efficiency, data-driven modeling approaches were studied to achieve fast and accurate flow prediction in the latest research[1][2]. The data driven flow modeling methods are divided into two types[3]. The one is based on reduced-order model (ROM), and the other is based on deep learning method. Both of them are widely used in flow prediction[4], flow mechanism analysis[5], control[6] and optimization[7].

Data-driven flow modeling methods based on ROM build a low-dimensional model of flow properties by using feature extraction methods, such as proper orthogonal decomposition (POD)[8] and dynamic mode decomposition (DMD)[9]. They have high modeling efficiency, and are widely applied in steady and unsteady flow modeling. Cao et al.[10] proposed a constrained ROM via POD to realize the prediction of steady hypersonic flows over 2D cylinder and 3D reentry vehicle. Sun et al.[11] proposed a non-intrusive ROM by combing POD and artificial neural network, and applied it to transonic flow prediction of variable geometries. In the work of Hijazi et al.[12], POD-Galerkin approach was used to build the ROM for turbulent flows. This ROM can be used to fast predict steady and unsteady flow field with high Reynolds. Yao et al.[13] employed POD to extract the modes of transonic unsteady flows. Additionally, they established an evolution model for POD mode coefficients using DMD with control (DMDc), enabling the prediction of transonic flow properties and flutter boundary.

In comparison with POD, DMD can obtain spatiotemporal coherent structures with frequency information and is superior in complex unsteady flow modeling, such as store separation of an aircraft[14], flow-induced vibration of tethered sphere[15], and flapping wings[16]. In the work of Zhang et al[17], DMD was initially employed to extract the modes of the flow past a 2D cycle cylinder. Subsequently, they proposed a flow sensing framework that combines DMD, long short-term memory (LSTM) network, and deep feedforward neural network to predict the unsteady flow based on sparse measurements. Liu et al.[18] employed DMD to analyze the convergence process of steady flow, and accelerate the convergence speed by 3-6 times. In the study conducted by Zhao et al.[19], DMD was used for mode analysis of the transonic flow over airfoils, which reveals the flow characteristics near buffet onset conditions. Jia et al.[20] proposed a hybrid ROM named DMD-PODX to improve modeling accuracy and robustness of unsteady flow.

Most of the previous flow modeling methods based on ROM were carried out with the linearized hypothesis of high-order systems. However, the flow in aerodynamics usually involves complex and strong nonlinear flow structures, such as shock wave and vortex. Thus it is challenging to ensure modeling accuracy in regions with nonlinear flow using data-driven flow modeling method base on ROM.

With wide applications of artificial intelligence in fluid mechanics[21][22], many data-driven flow modeling methods based on deep learning have been developed in recent years. Deep learning has powerful learning ability due to its complex network and advanced learning algorithms. Therefore, it has good potential in handling flow modeling problems with high dimension and strong nonlinear characteristics. In the work of Sun et al.[23], a deep learning framework based on multi-layer perceptron neural network was applied in transonic flow prediction over airfoils. Bhatnagar et al.[24] used convolutional neural network (CNN) to construct an approximation model that relates flow conditions and airfoil shapes to the resulting flow field. The architecture based on CNN was also applied in fast pressure distribution prediction, and it was verified that CNN is superior to feedforward neural network (FNN) in terms of accuracy and efficiency[25]. Peters et al.[26] raised the flow prediction models of fixed-wing store separation respectively using CNN and POD-based model. The results demonstrated that both models achieved high-precision prediction of surface distributed load, while the POD-based model required lower computational cost compared to the CNN model. Hu et al.[27] designed a new convolution operator with mesh resolution independence, and applied it in subsonic airfoil flow field prediction. Zuo et al.[4] proposed a deep learning architecture for flow field prediction based on CNN and multi-head perceptron neural network.

In addition, some new network architectures and modeling strategies have been adopted. In the work of Ma et al.[28], a flow modeling framework via residual neural network (Resnet) and POD was proposed to achieve flow field reconstruction. The results demonstrated that the Resnet is superior to FNN in terms of model accuracy and robustness. While this method is a non-intrusive ROM, which will lose accuracy in the process of reduced-order modeling. Lei et al.[29] raised an inverse design method of supercritical airfoil by combing generative adversarial network, CNN, and genetic algorithm, which can quickly obtain optimal airfoils. Thuerey et al.[30] introduced U-net into the steady flow field modeling, then they proposed a new deep learning framework named FlowDNN[31] by adding attention mechanisms, which can better extract physical information of flow field and enhance the flow prediction accuracy. Liu et al.[32] developed an enhanced hybrid deep neural network architecture for transonic unsteady flow prediction, which consists CNN and convolutional LSTM.

The above data-driven flow modeling methods based on deep learning are suitable for complex and nonlinear flow. However, due to high-dimensional modeling data, the modeling efficiency is much lower than that of ROM. In this study, to enhance the modeling accuracy while maintaining efficiency, we propose an accuracy-enhanced transonic flow modeling method by combing deep learning and ROM. Initially, the CNN-POD is used to build a prediction model mapping geometries to entire flow properties. Subsequently, the local flow regions that contain complex and nonlinear flow structures for enhanced modeling are identified through POD reconstruction. Next, the CNN is utilized to build a prediction model mapping geometries to local flow properties. Finally, the enhanced DNN achieves modeling and prediction of flow properties by fusing the CNN-POD and CNN models.

The following are the contributions of the present work:

1. A novel enhanced DNN model for flow prediction is proposed by fusing ROM and deep learning approaches, aiming to improve modeling accuracy of flow fields with strongly nonlinear flow structures.

2. A strategy to identify regions containing complex and nonlinear flow structures is developed using POD reconstruction.

3. The differences in accuracy, efficiency, and robustness between the enhanced DNN model and the CNN-POD and CNN models are evaluated by modeling different flow properties.

The remainder of the paper is organized as follows. Section 2 presents the basic methods, including the POD, and CNN. Section 3 describes the procedure of the proposed enhanced flow prediction method, and briefly introduces two other comparative methods. In Section 4, the proposed method is verified by predicting transonic flow over airfoils. Section 5 gives the conclusions of this paper.

## 2. Basic methods

### 2.1. Proper orthogonal decomposition for flow field

In this study, POD is used to extract modes from flow properties, such as pressure, velocity components, etc. The objective is a ROM which can achieve reduced-order and reconstruction of high-dimensional flow field data. The POD modeling process is described as follows.

A sampled snapshot sequence from steady flow properties with various shapes by computational fluid dynamics (CFD) simulation is described as $X = [X_1, X_2, \cdots, X_r]$, where $X_i \ (i=1,2,\ldots,r)$ is an $n$-dimensional column vector, $n$ is the number of mesh elements of the computational grid, and $r$ is the number of sampled shapes. Generally, $n \gg r$. Define the mean flow as $\bar{x}$, which represents the mean of $X$, and is calculated by

$$\bar{x} = \sum_{i=1}^{r} X_i \Big/ r \tag{1}$$

The standardized flow data $x = [x_1, x_2, \cdots, x_r]$ can be obtained by subtracting $\bar{x}$ from $X$. To reduce the memory consumption and improve the computational efficiency, the snapshot-POD[33] is adopted.

Define the correlation matrix of $x$ as $C \in \mathbb{R}^{r \times r}$.

$$C = x^\mathrm{T} x \tag{2}$$

Through the feature decomposition of $C$, the eigenvector matrix $Q = [q_1, q_2, \cdots, q_r] \in \mathbb{R}^{r \times r}$ and the eigenvalue matrix $\lambda = diag(\lambda_1, \lambda_2, \cdots, \lambda_r)$ ($\lambda_1 \geq \lambda_2 \geq \cdots \geq \lambda_r$) are obtained, which satisfies

$$CQ = \lambda Q \tag{3}$$

where eigenvalue $\lambda_i$ represents the energy of POD mode $\xi_i$, the energy proportion of $\xi_i$ is defined as $K_i = \lambda_i / \sum_{i=1}^{r} \lambda_i$, which is used to measure the contribution of $\xi_i$ in the flow field.

Then, the POD mode $\xi = [\xi_1, \xi_2, \cdots, \xi_r] \in \mathbb{R}^{n \times r}$ can be calculated by

$$\xi_i = x q_i / \sqrt{\lambda_i} \tag{4}$$

Define the POD mode coefficients corresponding to $x_i$ as $a_i$, which is calculated by

$$a = [a_1, a_2, \cdots, a_r] = \begin{bmatrix} a_1^1 & a_2^1 & \cdots & a_r^1 \\ a_1^2 & a_2^2 & \cdots & a_r^2 \\ \vdots & \vdots & \vdots & \vdots \\ a_1^r & a_2^r & \cdots & a_r^r \end{bmatrix} = \xi^\mathrm{T} x \tag{5}$$

We select the first $m$ modes to involve in the reduced-order modeling, while $m$ is determined by the error analysis of POD reconstruction. Correspondingly, the first $m$ coefficients of $a_i$ are retained: $a_i = [a_i^1, a_i^2, \cdots, a_i^m]^\mathrm{T}$. $X^i$ can be approximately expressed as

$$X_i \approx \xi a_i + \bar{x} \tag{6}$$

### 2.2. Convolutional neural network for regression

As a primary and classical method of deep learning approaches, CNN can extract the relevant feature from high-dimensional data[34] and is widely applied in fluid mechanics recently[35][36]. In this work, CNN is used to construct a regression model from geometries to flow properties, as is shown in Fig. 1. The architecture consists of image input layer, several sets of convolutional and pooling layers, fully connected layer, and regression output layer. In the image input layer, the geometries are characterized by signed distance function (SDF) sampled on a Cartesian grid, which can be seen in section 3.1. In the convolution layers, BatchNorm is used to improve the convergence efficiency and robustness of the training process, while Relu (Linear rectification function) is selected as the activation function. In the pooling layer, average pooling is adopted. In this paper, the pooling filter size is 2×2, and the stride is 2. The fully connected layer is used to fuse the feature matrices obtained from the last convolution and pooling layers, and flatten them into a one-dimensional vector. In the regression output layer, the flow properties can refer to the local and entire flow field, or be represented by the mode coefficients of flow field, while a loss function is defined by the mean square error (MSE), which is expressed by

$$L_2 = \sum (y' - y)^2 \tag{7}$$

where $y$ represents the real response and $y'$ represents the predicted response for validation set during the network learning process.

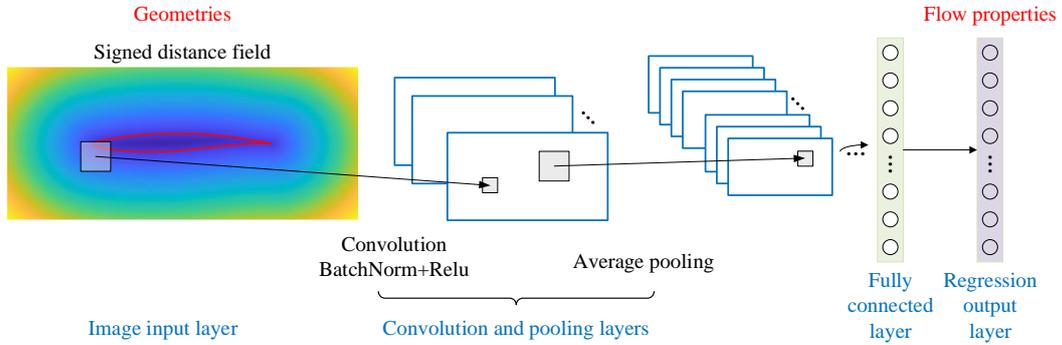

Fig. 1 The regression CNN architecture for regression, which maps geometries to flow properties

## 3. Flow prediction methods

In this section, an enhanced DNN model is proposed for transonic flow prediction, which combines ROM and deep learning. The enhanced DNN model can effectively improve modeling accuracy. In order to thoroughly evaluate the performance of the proposed flow prediction method, two additional commonly employed flow prediction methods, namely the CNN-POD model and the CNN model, are introduced for comparison.

The above three methods are used to map the geometries represented by SDF $\boldsymbol{D} = [\boldsymbol{D}_1, \boldsymbol{D}_2, \cdots, \boldsymbol{D}_r]$ to flow field $\boldsymbol{X} = [\boldsymbol{X}_1, \boldsymbol{X}_2, \cdots, \boldsymbol{X}_r] \in \mathbb{R}^{n \times r}$, where $\boldsymbol{D}_i \in \mathbb{R}^{W \times H}$ $(i = 1, 2, \cdots r)$ represents the $i$th geometry using SDF representation. $\boldsymbol{X}$ is defined in the computational domain (See section 3.1) and represents the flow field data with $n$-dimensional of $r$ samples in the training set.

### 3.1. Geometry representation and coordinate transformation

Geometry can be represented in various ways, including boundaries, geometric parameters, and point clouds[37]. In this paper, we adopt SDF to achieve a generic input representation for CNN that can handle various shapes. SDF characterizes geometric shapes by representing the minimum distance from each point in Cartesian grid to the surface of geometry. The signed distance can be expressed by

$$D(p) = \begin{cases} d(p,\partial\Omega) & p \notin \Omega \\ 0 & p \in \partial\Omega \\ -d(p,\partial\Omega) & p \in \Omega \end{cases} \quad (8)$$

where $p$ is the point in the Cartesian grid, $D(p)$ represents the signed distance located on $p$; $\Omega$ donates the geometry, while $\partial\Omega$ donates the surface of geometry; Define $s$ is the point located on the $\partial\Omega$. $d(p,\partial\Omega) = \min_{s\in\partial\Omega}|p-s|$ measures the minimum distance from $p$ to $\partial\Omega$.

The validity of SDF in intelligence fluid mechanics has been fully verified[38][39]. Fig. 2 illustrates the SDF contour plot for the RAE2822 airfoil with a 128×64 Cartesian grid. The display region is within $-0.5 \le x/c \le 1.5$ and $-0.5 \le y/c \le 0.5$, where $c$ is the chord length of airfoil.

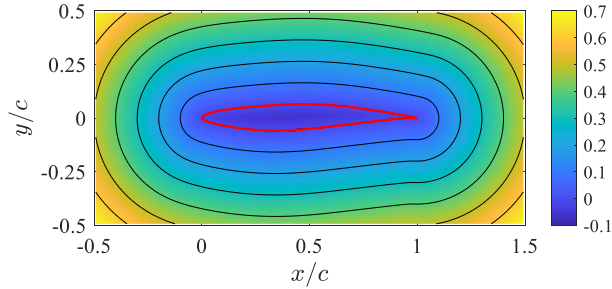

**Fig. 2 The SDF representation of the RAE2822 airfoil with a 128×64 Cartesian grid.**

The direct use of POD and CNN is hindered in flow problems with strongly nonlinear flow structures due to the prevalent use of non-uniform meshes in numerical simulations. The projection of the flow field onto a uniform Cartesian mesh results in insufficient information about the turbulent flow field in the near-wall region. To address this limitation, we employ a transformation technique to map the flow field from the non-uniform physical domain to the uniform computational domain[40]. The relationship between the physical domain and computational domain is shown in Fig. 3. Fig. 4 shows the difference between the physical domain and computational domain for the pressure field. It can be seen that the computational domain enables the preservation and visualization of the main flow structure in the near-wall region.

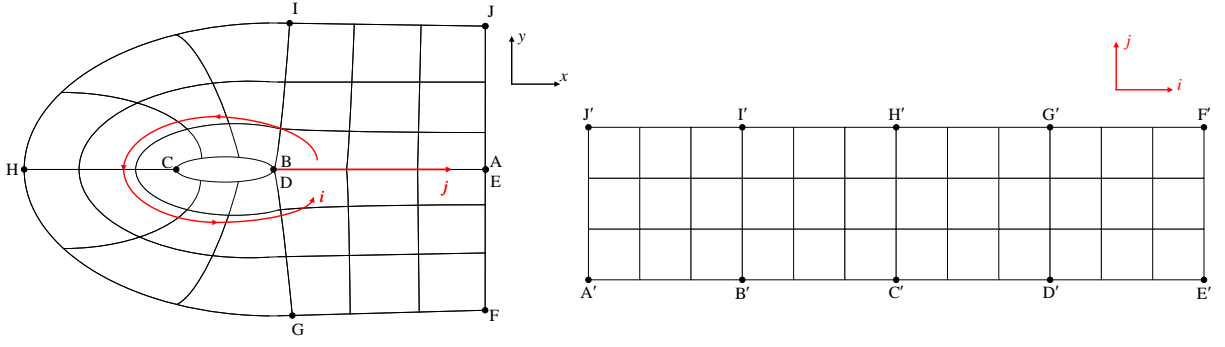

**Fig. 3 The physical domain (left) and the computational domain (right).**

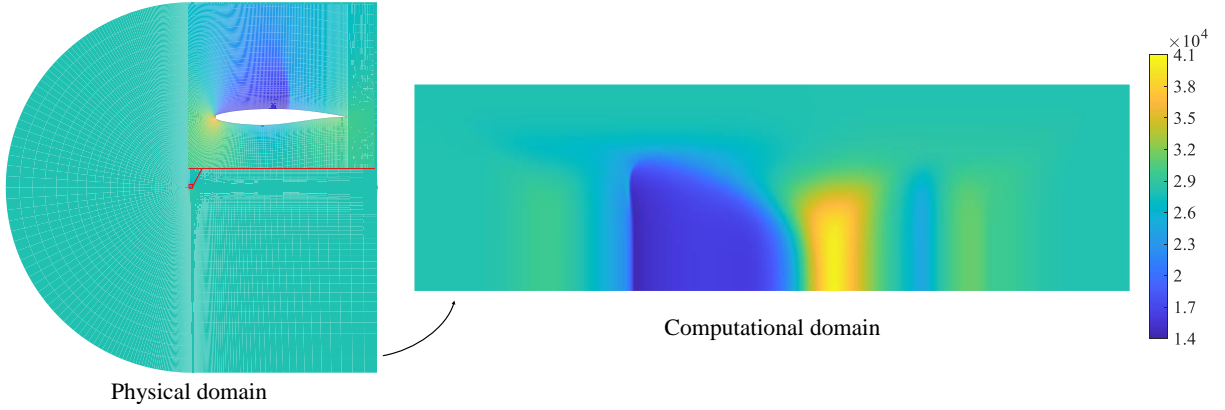

**Fig. 4 The physical domain and computational domain for the pressure field of the RAE2822 airfoil, where the unit is Pa.**

### 3.2. CNN-POD

The CNN-POD model is a data-driven flow modeling method based on ROM, and the schematic diagram of flow modeling and predicting are shown in Fig. 5. In the modeling stage, a POD model $\psi_{\text{POD}}$ is established first, which can be expressed by

$$[\bar{x},\xi,a] = \psi_{\text{POD}}(X) \tag{9}$$

where $\bar{x}$ is the mean flow, $\xi = [\xi_1, \xi_2, \cdots, \xi_m] \in \mathbb{R}^{n \times m}$ are the selected flow modes, $a = [a_1, a_2, \cdots, a_r] \in \mathbb{R}^{m \times r}$ are the mode coefficients. Thus, the $n$-dimensional flow field data $X$ can be approximated by $X \approx \xi a + \bar{x}$. Then, a CNN model $\psi_{\text{CNN}}$ is used to map the geometries to the mode coefficients, that is

$$a = \psi_{\text{CNN}}(D) \tag{10}$$

The dimensions of the input and output for $\psi_{\text{CNN}}$ are $W \times H$ and $m$, respectively. The flow prediction method based on CNN-POD is built by combing $\psi_{\text{POD}}$ and $\psi_{\text{CNN}}$.

In the predicting stage, a new geometry represented by SDF $D' \in \mathbb{R}^{W \times H}$ is provided as input. Then its corresponding mode coefficients $a'$ are predicted by the trained CNN model $\psi_{\text{CNN}}$.

$$a' = \psi_{\text{CNN}}(D') \tag{11}$$

Thereafter, the reconstructed flow field can be expressed by

$$X' \approx \xi \psi_{\text{CNN}}(D') + \bar{x} \tag{12}$$

Finally, the predicted flow field is obtained by converting from the computational domain to the physical domain.

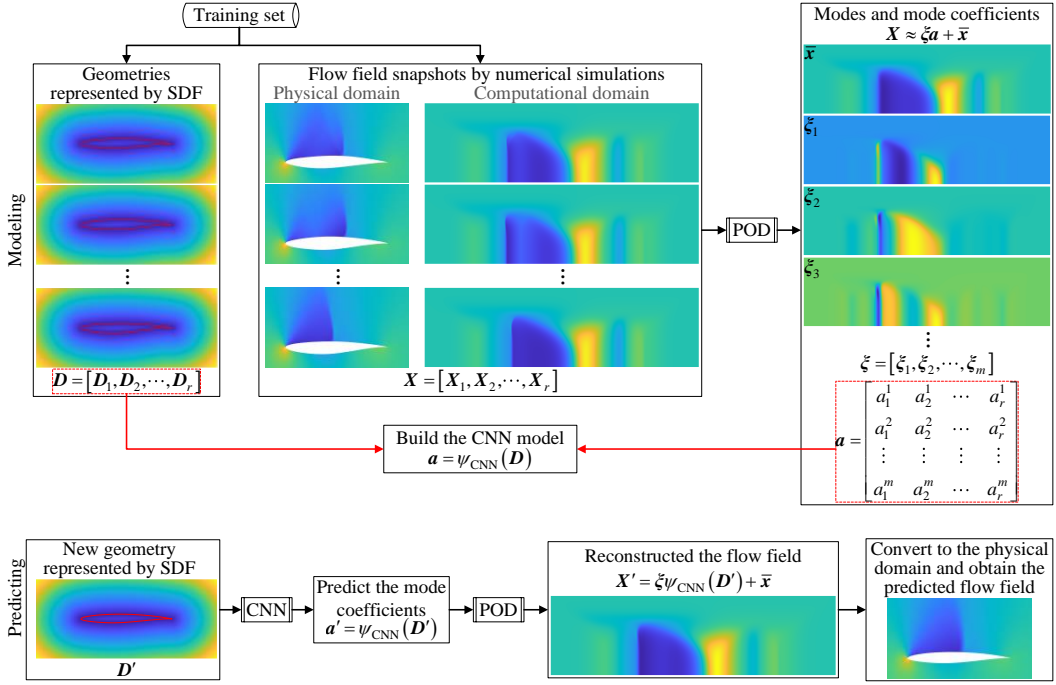

**Fig. 5 The schematic diagram of flow modeling and predicting using the CNN-POD model**

### 3.3. CNN

The CNN model is a data-driven flow modeling method based on deep learning, and the flow modeling and predicting processes are shown in Fig. 6. In the modeling stage, a CNN model $\psi_{\text{CNN}}$ is established to map geometries to flow field directly, that is

$$X = \psi_{\text{CNN}}(D) \tag{13}$$

The input dimension for $\psi_{\text{CNN}}$ here is $W \times H$, which matches the input dimension mentioned in section 3.2. Meanwhile, the output dimension for $\psi_{\text{CNN}}$ here is $m$, which is usually significantly smaller than the output dimension mentioned in section 3.2. Reducing the output dimension can result in fewer network parameters, thus enhancing the training efficiency of the network.

In the predicting stage, a new geometry represented by SDF $D'$ is given. The predicted flow field can be obtained through the trained CNN model.

$$X' = \psi_{\text{CNN}}(D') \tag{14}$$

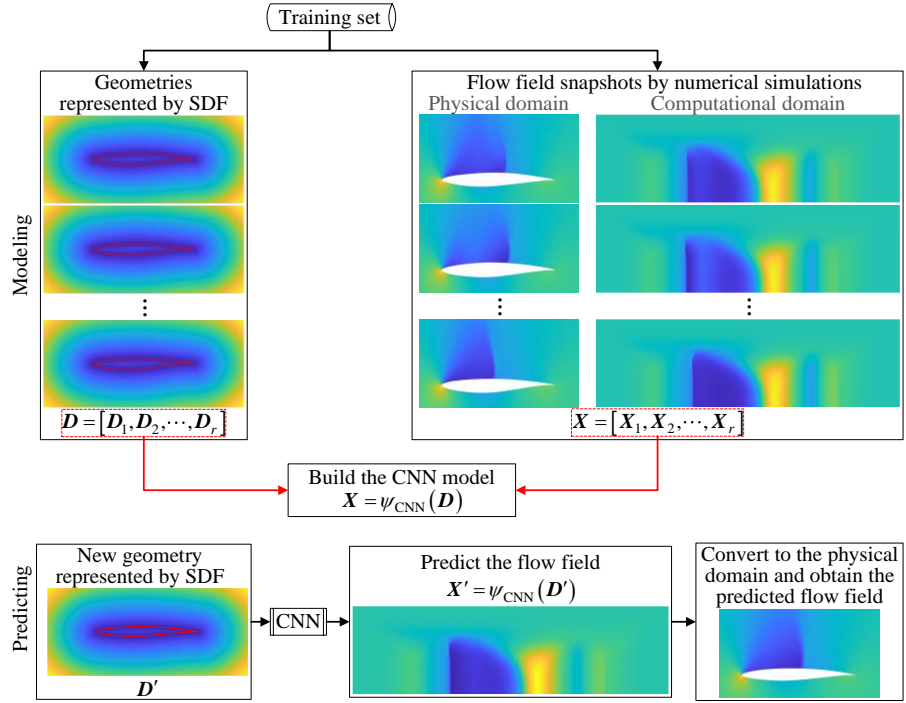

**Fig. 6 The schematic diagram of flow modeling and predicting using the CNN model**

### 3.4. Enhanced DNN

The enhanced DNN is proposed as a flow modeling method, which combines CNN-POD and CNN model. The proposed method integrates the advantages of both CNN and CNN-POD methods, and provides performance improvements in modeling accuracy and efficiency. Fig. 7 and Fig. 8 illustrates the schematic diagram of flow modeling and predicting using the enhanced DNN model, respectively.

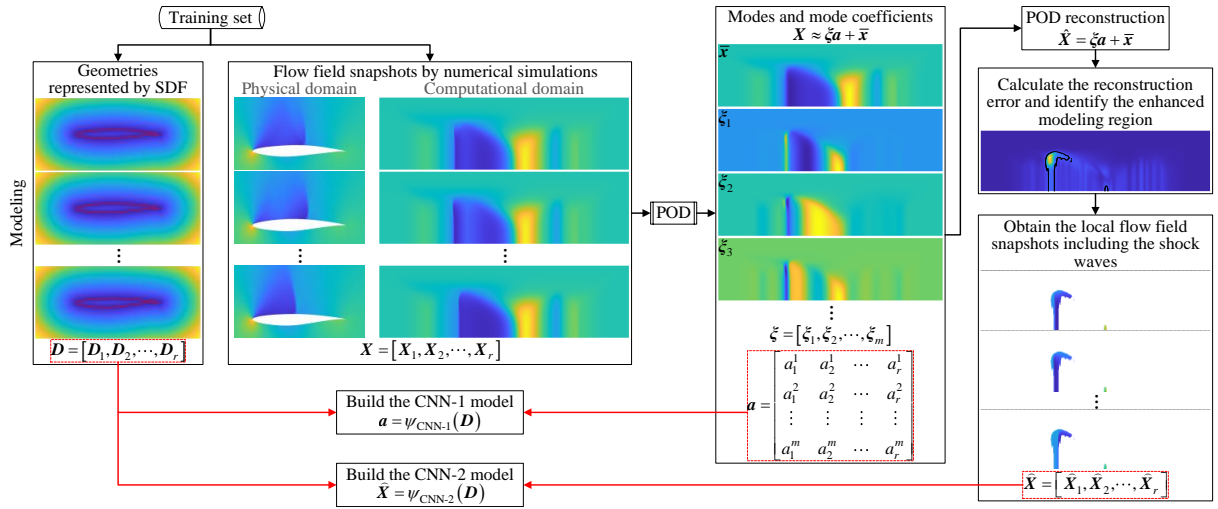

**Fig. 7 The schematic diagram of flow modeling using the enhanced DNN model**

In the modeling stage, the CNN-POD model is used to build the map from the geometries to the entire flow field first, which is the same as in section 3.2 and expressed by

$$\begin{cases} [\bar{x}, \xi, a] = \psi_{POD}(X) \\ a = \psi_{CNN\text{-}1}(D) \end{cases} \tag{15}$$

where $\psi_{POD}$ and $\psi_{CNN\text{-}1}$ are part of the CNN-POD model. Then, the flow field can be reconstructed by

$$\hat{X} = \xi a + \bar{x} \tag{16}$$

Furthermore, the reconstructed error $E = [E_1, E_2, \cdots, E_n] \in \mathbb{R}^{n \times 1}$ is calculated.

$$E = \frac{1}{r} \sum_{i=1}^{r} |\hat{X}_i - X_i| \tag{17}$$

Then, we define a threshold index $\sigma$ to identify the regions with poorer reconstructed performance and obtain the local flow field $\widehat{X} = [\widehat{X}_1, \widehat{X}_2, \cdots, \widehat{X}_r] \in \mathbb{R}^{b \times r}$ for enhanced modeling.

$$\begin{aligned} \widehat{X}_i &= [X_{c_1,i}, X_{c_2,i}, \cdots, X_{c_b,i}]^T \\ \{c_1, c_2, \cdots, c_b\} &= \{j \mid j = 1, 2, \cdots, n,\ E_j \geq \sigma\} \end{aligned} \tag{18}$$

where $b$ is the data dimension of the local flow field, which satisfies $b \leq n$. Typically, the identified local flow field consists of strongly nonlinear flow structures that are challenging to model using POD.

Thereafter, another CNN model $\psi_{CNN\text{-}2}$ is adopted to build the map from geometries $D$ to the local flow field $\widehat{X}$.

$$\widehat{X} = \psi_{CNN\text{-}2}(D) \tag{19}$$

In comparison with the CNN model $\psi_{CNN}$ in section 3.3, $\psi_{CNN\text{-}2}$ concentrates on the local flow field, which is characterized by shock waves and poses a challenge accurate modeling using the POD model. It is obvious that $\psi_{CNN\text{-}2}$ outperforms $\psi_{CNN}$ in terms of modeling accuracy and efficiency. This can be attributed to the fact that $\psi_{CNN\text{-}2}$ requires less flow field data for modeling, which allows it to achieve better modeling accuracy with less modeling time.

So far, the enhanced DNN model has been built by combining the CNN-POD model for entire flow field and CNN model for local flow field. It is worth noting that the threshold index $\sigma$, is a critical factor that influences the performance of the proposed enhanced model. The discussion on the impact of selecting different thresholds will be presented in section 4.3.

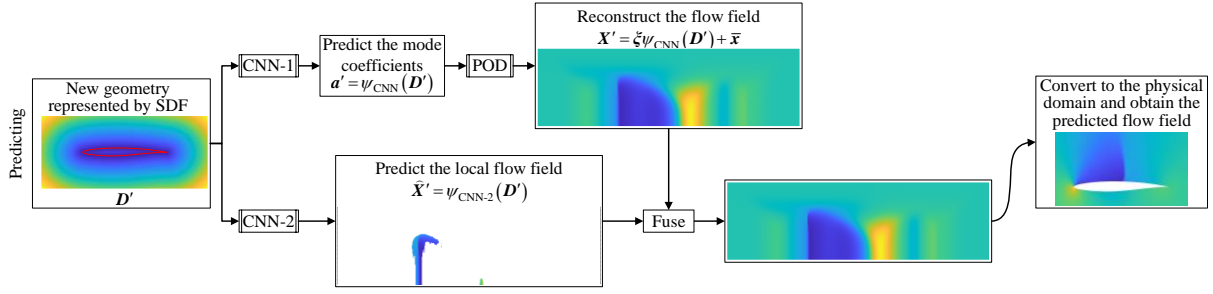

**Fig. 8 The schematic diagram of flow predicting using the enhanced DNN model**

In the predicting stage, a new geometry represented by SDF $D'$ is given. First, the predicted entire flow field $X' = [X'_1, X'_2, \cdots, X'_n]^T$ is solved by $\psi_{CNN-1}$ and $\psi_{POD}$, which is same as the CNN-POD model in section 3.2.

At the same time, the predicted local flow field $\widehat{X}' = [\widehat{X}'_{c_1}, \widehat{X}'_{c_2}, \cdots, \widehat{X}'_{c_b}]^T$ is solved by $\psi_{CNN-2}$, that is

$$\widehat{X}' = \psi_{CNN-2}(D') \tag{20}$$

Furthermore, the predicted flow field $X'' = [X''_1, X''_2, \cdots, X''_n]^T$ can be obtained by fusing $X'$ and $\widehat{X}'$.

$$X''_i = \begin{cases} X'_i & i \notin \{c_1, c_2, \cdots c_b\} \\ \widehat{X}'_i & i \in \{c_1, c_2, \cdots c_b\} \end{cases} \tag{21}$$

Finally, the predicted flow field is obtained by converting from the computational domain to the physical domain.

## 4. Test case

### 4.1. Physical model

The proposed enhanced DNN method is validated by transonic steady flow prediction of airfoil. In this case, the flow conditions are $Ma = 0.734$, $\alpha = 2.79°$, $Re = 6.5 \times 10^6$ and $c = 1$. The airfoil RAE2822 is adopted as the initial baseline. The class shape function transformation (CST)[41] approach is adopted to parameterize airfoil and product new various shapes. Here, the 6th-order CST polynomials with 14 parameters, 7 parameters for upper and lower airfoil surface, respectively, is employed to describe the airfoil shapes. Then 1200 samples of various airfoils are generated by Latin hypercube sample method[25]. Fig. 9 shows all the airfoils, which including 900 training samples and 300 testing samples.

The density-based compressible RANS (Reynolds-averaged Navier-Stokes) solver with SST k-omega turbulence model is specified to achieve accurate numerical simulation. The computational mesh and accuracy verification of the solver are shown in Appendix A. Through numerical simulations, the flow field data of airfoils including pressure, velocity components are obtained.

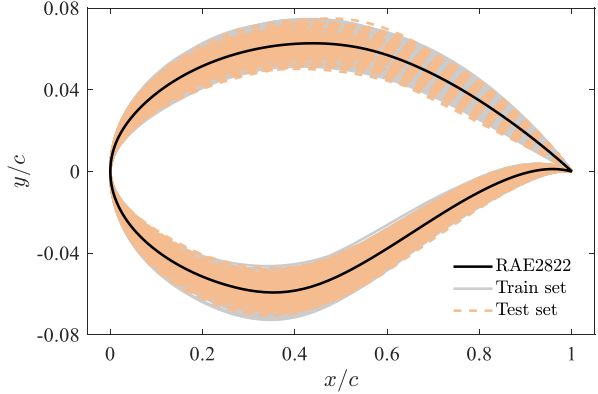

**Fig. 9 The airfoil database including the baseline airfoil, train set, and test set.**

### 4.2. Modeling and training

The airfoil geometries are represented by SDF with a 128×64 Cartesian grid, and serve as the input for the flow modeling model. The flow field data is transformed from the non-uniform physical domain into the uniform computational domain, and utilized for POD and CNN model.

To ascertain the number of modes involved in the POD modeling process, the reconstructed error $e$ is defined as follows and employed to assess the reconstruction accuracy of the POD model.

$$e = \frac{1}{r}\sum_{i=1}^{r}\left(\frac{1}{n}\|\hat{\boldsymbol{y}}_i - \boldsymbol{y}_i\|_1\right) \tag{22}$$

where $r$ is the number of the train samples; $\hat{\boldsymbol{y}}_i$ and $\boldsymbol{y}_i$ are respectively the reconstructed and real flow field of the $i$th train sample; $n$ is the dimension of $\boldsymbol{y}_i$. $\|\cdot\|_1$ represents the 1-norm.

For the POD modeling of pressure field, the energy proportion and reconstructed error are shown in Fig. 10. It reveals that the initial modes are dominant modes with high energy proportions, consequently leading to a rapid decline in reconstructed error. The first 5 POD modes are is shown in Fig. 11. Remarkably, the cumulative energy proportion of the first 30 modes reaches 99.14%, while the reconstructed error drops to 21.9 Pa. This observation indicates that the transonic flow characteristics over airfoils mainly reside within the first 30 modes, and the POD model utilizing these 30 modes demonstrates good dimensionality reduction performance. Further, increment in the number of modes can not enhance the reconstruction accuracy of the flow field. Therefore, we selected 30 modes for constructing the POD model in this study.

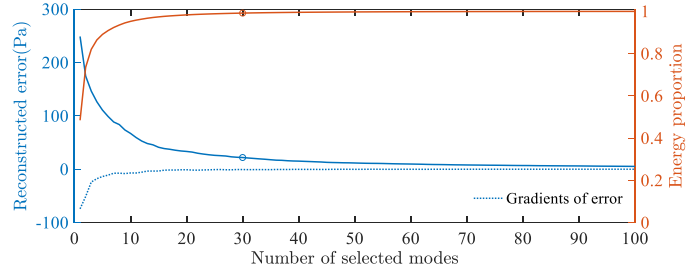

**Fig. 10 The reconstructed error and energy proportion of the POD model for the pressure field.**

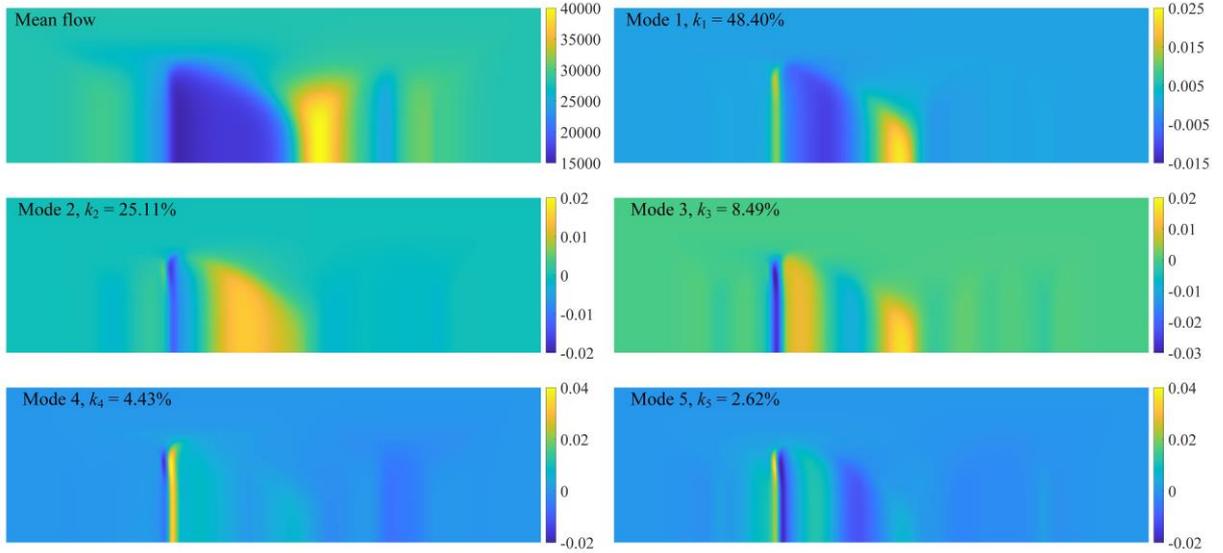

**Fig. 11 Dominant POD modes of pressure field on the computational domain, where the unit is Pa.**

In the process of modeling enhanced DNN, the local flow filed snapshots are calculated by POD reconstruction analysis. As is shown in Fig. 12, the enhanced modeling regions of the pressure field are identified with the condition of $\sigma = 150$ Pa and $\sigma = 100$ Pa. In comparison with the computational domain in Fig. 4, it can be seen that the identified regions coincide with the complex flow regions containing shock waves. The data dimension of the identified local flow field depends on the value of $\sigma$. In this particular case, when $\sigma = 100$ Pa, the data dimension of the local flow field is 1921, which accounts for merely 2.6% of the overall data dimension of the entire flow field.

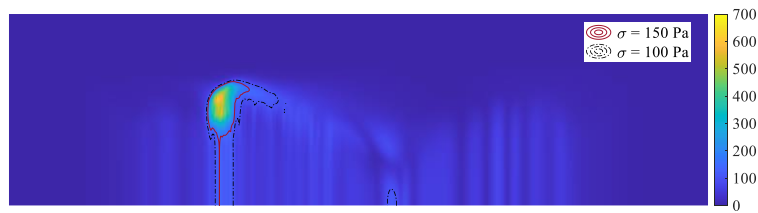

**Fig. 12 The mean reconstructed error of the pressure field on the computational domain.**

Furthermore, in order to achieve optimal modeling accuracy and efficiency for diverse flow prediction models, we carefully select suitable CNN model architectures and hyperparameters. The selection of architectures and hyperparameters for CNN models utilized in three flow prediction models is intricately connected to the sizes of the input and output data. Since the inputs for all three models consist of consistent geometries represented by SDF, a higher output dimension in the model corresponds to an increase in model hyperparameters. As an example of pressure flow field modeling, Table 1 shows cases the architectures and hyperparameters of four distinct CNN models in different flow prediction methods. The models encompass CNN-POD (utilizing Net-1), CNN (utilizing Net-2), and enhanced DNN model with different thresholds (utilizing Net-3 and Net-4). For all the CNN models utilized in this study, the initial learning rate is set to 0.001, the batch size is set to 128, and a total of 5000 epochs are conducted for training. The training was performed on a PC equipped with a GPU (RTX3090Ti).

Table 1 The description of each layer and parameters of various networks. (In the 1st Conv-Layer of Net-1, 5×5 indicates the convolutional filter size, 64 indicates the number of convolutional filters. In the fully connected layer of Net-1, 30×512 indicates the size of weights parameters.)

| CNN | Net-1 | Net-2 | Net-3 | Net-4 |
|---|---|---|---|---|
| Image input layer | 64×128×1 | 64×128×1 | 64×128×1 | 64×128×1 |
| Filters of 1st Conv-layer | 5×5, 64 | 5×5, 64 | 5×5, 64 | 5×5, 128 |
| Filters of 2st Conv-layer | 5×5, 64 | 5×5, 128 | 5×5, 64 | 5×5, 128 |
| Filters of 3st Conv-layer | 5×5, 128 | 5×5, 256 | 5×5, 128 | 5×5, 256 |
| Filters of 4st Conv-layer | 3×3, 128 | 3×3, 512 | 3×3, 128 | 3×3, 256 |
| Filters of 5st Conv-layer | 3×3, 256 | 3×3, 1024 | 3×3, 256 | 3×3, 512 |
| Filters of 6st Conv-layer | 3×3, 256 | 3×3, 2048 | 3×3, 256 | 3×3, 512 |
| Dropout layer | Dropout rate = 0.2 | | | |
| Fully connected layer | 30×512 | 73904×4096 | 780×512 | 1921×1024 |
| Regression output layer | 30 | 73904 | 780 | 1921 |
| All parameters | 1.3M | 328.5M | 1.7M | 7.3M |

### 4.3. Prediction of pressure

In this section, three models for pressure prediction are respectively built by the CNN-POD, CNN, and enhanced DNN. Fig. 13 illustrates the error contours of predicted pressure. It is evident that the CNN-POD model shows the highest prediction error among all the predicted shapes, especially in the region surrounding the shock waves. The CNN model performs better than the CNN-POD model in terms of accuracy but still demonstrates poorer prediction performance in the region around the shock waves. The enhanced DNN model demonstrates superior prediction performance owing to its enhanced modeling strategy. The predicted error in the region surrounding the shock waves is further reduced compared to the CNN model. However, when $\sigma =150$ Pa, it appears significant error in the region near the airfoil surface because this region was not fully identified. On the other hand, when $\sigma =100$ Pa, the regions encompassing the entire shock wave can be accurately identified. Consequently, the prediction accuracy can be further improved compared to the enhanced DNN model with $\sigma =150$ Pa.

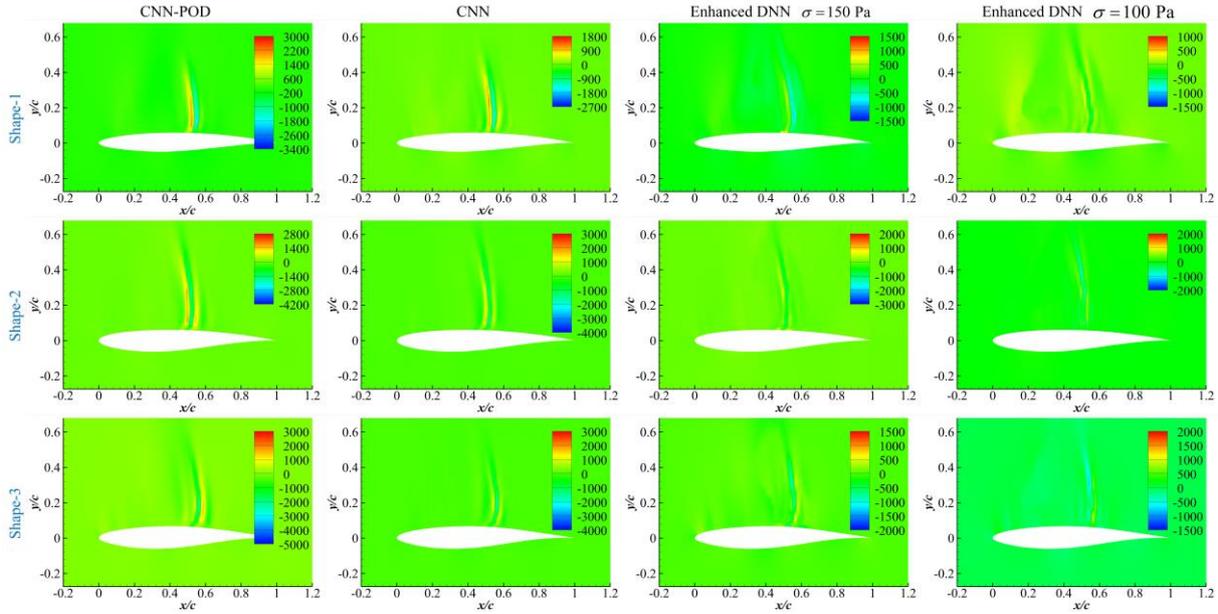

**Fig. 13 The prediction error of pressure field by different methods for three airfoils, where the unit is Pa.**

Fig. 14 illustrates the predicted pressure fields by three methods. The pressure isolines predicted by the CNN-POD model in the vicinity of the shock waves exhibit lower density than the reference isolines, indicating a weaker intensity of the predicted shock wave. The intensity of the shock waves predicted by the CCN model shows closer agreement with the reference values compared to the CNN-POD model, indicating better accuracy in directly modeling the region with shock waves. The pressure isolines predicted by the enhanced DNN model exhibit the closest agreement with the reference isolines, accurately predicting the intensity of the shock waves.

In addition, in order to investigate the accuracy of the method in predicting the surface pressure coefficient, we present a comparison of the pressure coefficient distribution over the surface of the airfoil between the predicted and reference results in Fig. 15. It is evident that all methods demonstrate good accuracy in predicting the pressure coefficients on the lower surface. However, the CNN-POD model shows noticeable differences in the prediction of the pressure coefficient on the upper surface, particularly in the region around the shock wave. In contrast, both the CNN and enhanced DNN models exhibit good agreement between the predicted pressure coefficients on the upper surface of the airfoil and the reference ones.

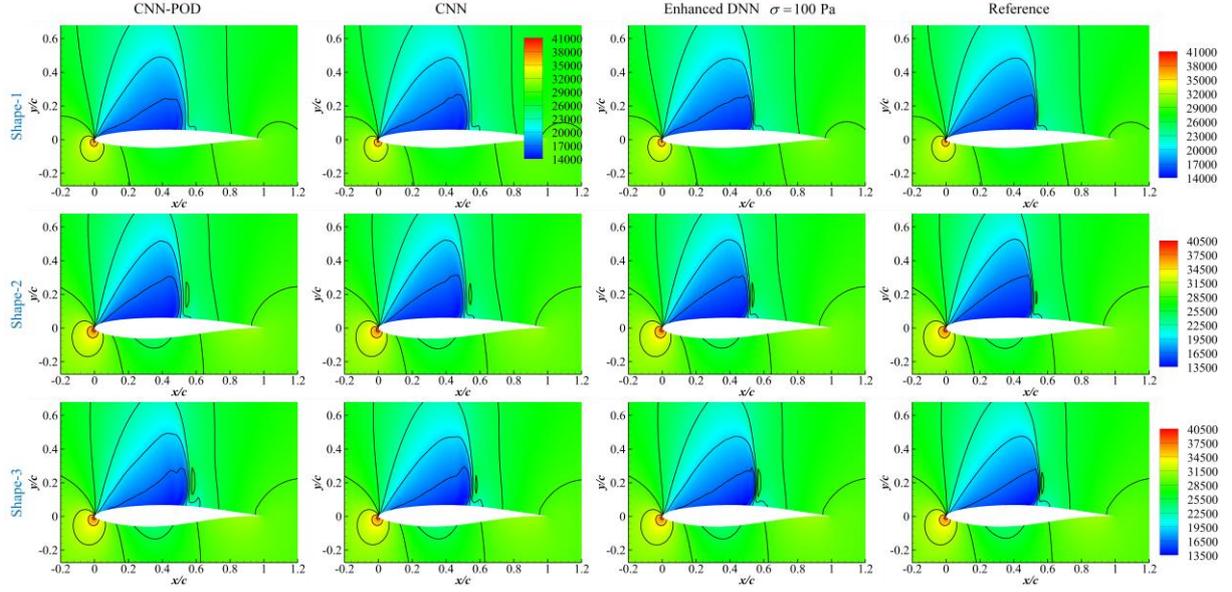

**Fig. 14 The forecasted results of pressure field by different methods for three airfoils, where the unit is Pa.**

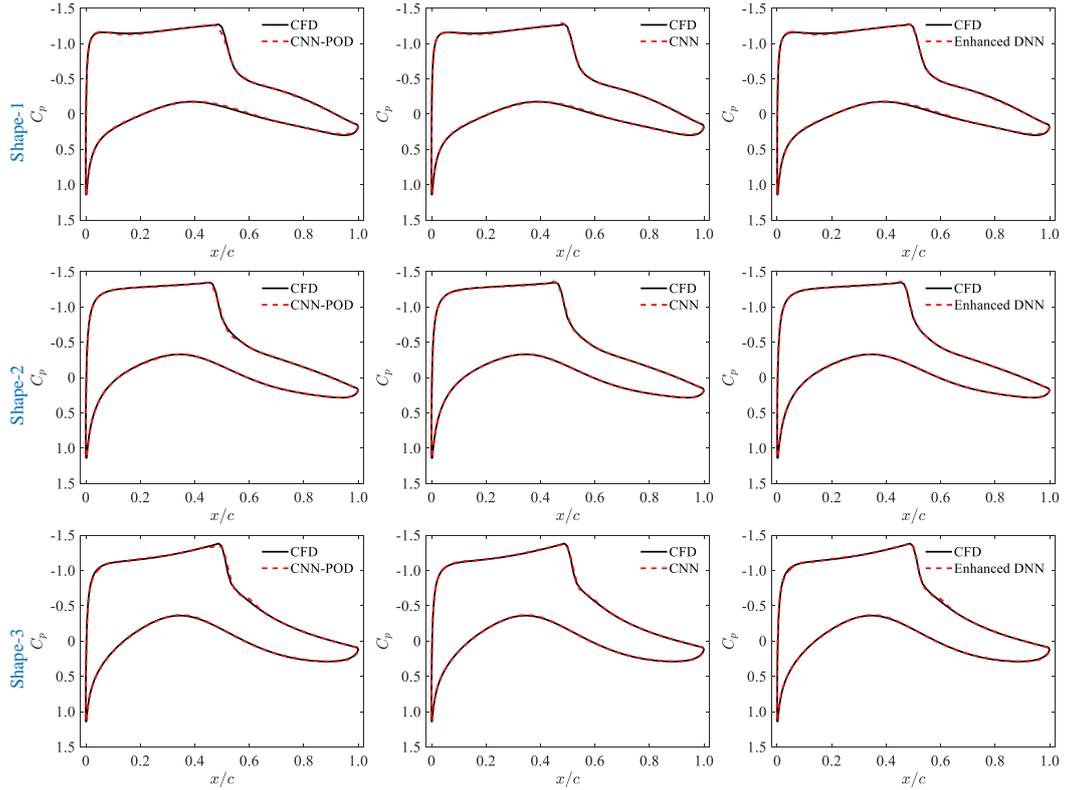

**Fig. 15 The forecasted results of surface pressure coefficient distribution by different methods for three airfoils.**

To ascertain the performance of different prediction methods specifically, the prediction error $E$ is formulated in a similar manner to the definition of the reconstructed error in Eq. (22), as follows.

$$E = \frac{1}{k} \sum_{j=1}^{k} \left( \frac{1}{n} \left\| \mathbf{y}'_j - \mathbf{y}_j \right\|_1 \right) \tag{23}$$

where $k$ is the number of the test samples. $\mathbf{y}'_j$ and $\mathbf{y}_j$ are respectively the predicted and real flow field of the $j$th test sample.

The modeling parameters and performance indicators for predicting pressure field using various methods are shown in Table 2, where, $T$ represents the training time of the CNN of each model. $E$-entire is the prediction error of the entire flow field. $E$-local represents the prediction error of the local flow field with the threshold $\sigma = 100$ Pa, which quantifies the modeling accuracy of the region containing shock waves.

It is evident that the enhanced DNN model with $\sigma = 100$ Pa exhibits the highest modeling accuracy in the region containing the shock wave. The prediction error is reduced by approximately 35.91% and 18.44% in comparison with the CNN-POD and CNN models, respectively. Additionally, the proposed method also demonstrates superior performance in predicting the entire flow field compared to other methods. The modeling time of enhanced DNN is 65.29% shorter than that of the CNN model. Additionally, the modeling parameter size of the enhanced DNN model is nearly two orders of magnitude smaller. These findings indicate that the enhanced modeling strategy has significantly improved both the modeling accuracy and efficiency when compared to the CNN model.

In addition, for the enhanced DNN model with $\sigma = 100$ Pa, the prediction error of the local flow field is 18.05% smaller compared to the enhanced DNN model with $\sigma = 150$ Pa. This indicates that the setting an appropriate threshold can effectively identify regions with poor POD modeling accuracy and improve the accuracy of the enhanced DNN model.

Table 2 The modeling parameters and performance indicators for pressure field prediction by different methods.

| Model | Input dimension | Output dimension | Parameters (M) | T (min) | $E$-entire (Pa) | $E$-local (Pa) |
|---|---|---|---|---|---|---|
| CNN-POD | 64×128 | 30 | 1.3 | 20 | 42.7 | 419.61 |
| CNN | 64×128 | 73904 | 328.5 | 170 | 39.06 | 329.72 |
| Enhanced DNN-1 ($\sigma$ =150 Pa) | 64×128 | 30+780 | 1.3+1.7 | 20+22 | 39.87 | 328.18 |
| Enhanced DNN-2 ($\sigma$ =100 Pa) | 64×128 | 30+1921 | 1.3+7.3 | 20+39 | 38.80 | 268.93 |

In order to further assess the robustness of the different prediction methods, all models are repeatedly trained 20 times and the prediction results are shown in Fig. 16. The CNN model exhibits a larger error distribution range in comparison with the CNN-POD and enhanced DNN models. The increased complexity of the CNN architecture for the entire flow field, compared to that for the mode coefficients and local flow field, is the reason behind this. Overall, the enhanced DNN model achieves the highest accuracy in predicting the flow field containing nonlinear flow structures while maintaining stable robustness.

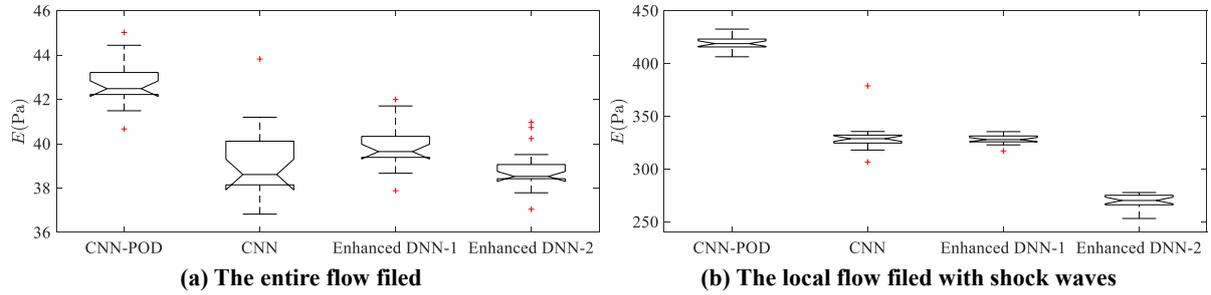

(a) The entire flow filed  (b) The local flow filed with shock waves

Fig. 16 Box plots for the prediction error of pressure by different methods of 20 runs.

### 4.4. Prediction of velocity

In this section, the prediction models for streamwise velocity and crossflow velocity are respectively built.

Fig. 17 and Fig. 18 illustrate the error contours and predicted results of streamwise velocity fields, respectively. Similar to the predicted results for the pressure field, the CNN-POD and CNN models show significant prediction error for all the predicted shapes, especially in the region around the shock wave. The streamwise velocity isolines predicted in the region around the shock wave exhibit lower density compared to the reference isolines, indicating a weaker intensity of the predicted shock wave. While the enhanced DNN model achieve the highest prediction accuracy, and the predicted streamwise velocity isolines are consistent with the reference isolines.

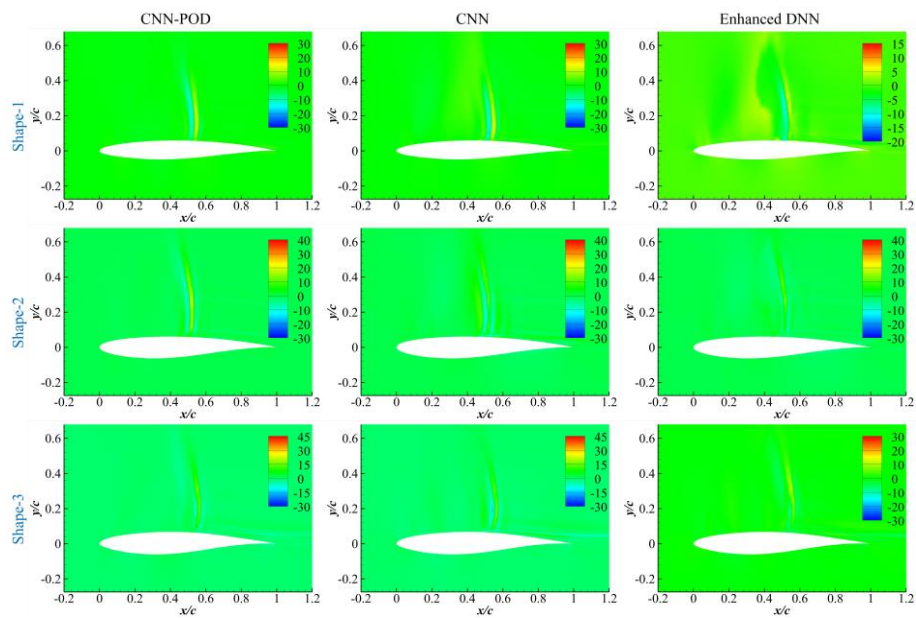

Fig. 17 The prediction error of streamwise velocity by different methods for three airfoils, where the unit is m/s.

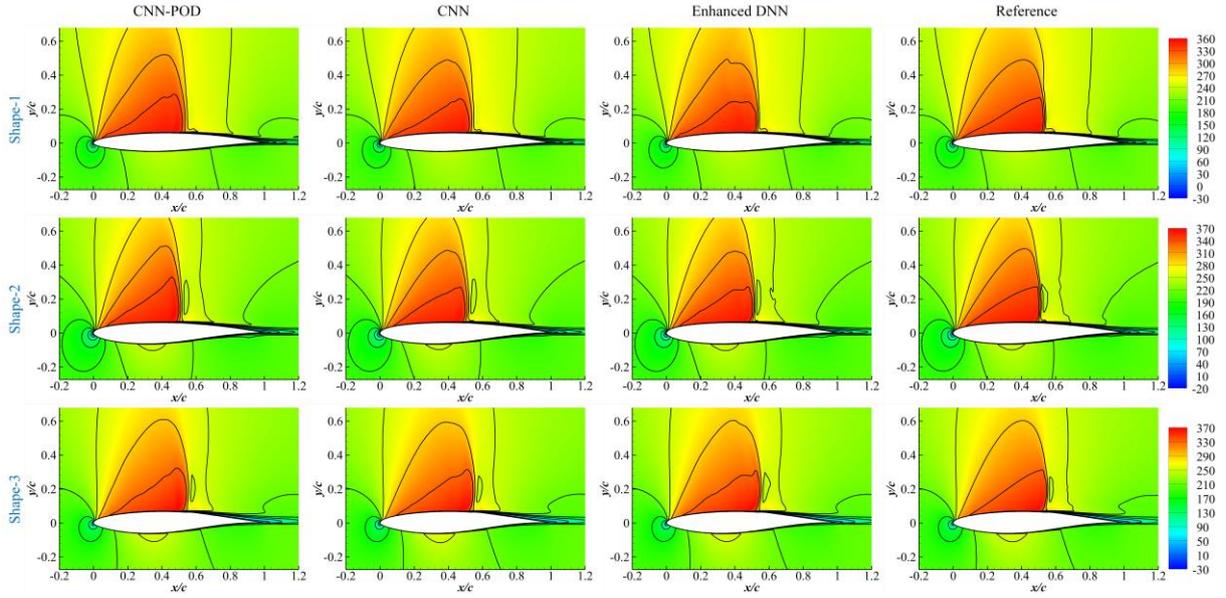

**Fig. 18 The forecasted results of streamwise velocity by different methods for three airfoils, where the unit is m/s.**

The modeling parameters and performance indicators for streamwise velocity prediction using different methods are shown in Table 3. It can be seen that the enhanced DNN model has best modeling accuracy in the region containing the shock wave. The prediction error is reduced by about 31.77% and 13.67% in comparison with CNN-POD and CNN models, respectively. The entire flow field prediction performance of the proposed method is close to CNN model and superior to CNN-POD model. Furthermore, the modeling time of enhanced DNN is 68.18% shorter than that of the CNN model, and the modeling parameter size of enhanced DNN model is 97.41% smaller than that of the CNN model. These results indicate that the proposed method has significantly improved both the modeling accuracy and efficiency compared to the CNN model.

**Table 3 The modeling parameters and performance indicators for the prediction of streamwise velocity by different methods**

| Model | Input dimension | Output dimension | Parameters (M) | T (min) | $E$-entire (m/s) | $E$-local (m/s) |
|---|---|---|---|---|---|---|
| CNN-POD | 64×128 | 30 | 1.3 | 20 | 0.67 | 6.20 |
| CNN | 64×128 | 73904 | 328.5 | 176 | 0.61 | 4.90 |
| Enhanced DNN | 64×128 | 30+1855 | 1.3+7.2 | 20+36 | 0.62 | 4.23 |

All models are repeatedly trained 20 times, and the prediction results are presented in Fig. 19. In terms of modeling the entire flow field, the CNN model shows the highest potential for prediction. However, its error distribution range is larger compared to the CNN-POD and enhanced DNN models. On the other hand, for modeling of the local flow field, the enhanced DNN model exhibits stable performance and achieves the highest prediction accuracy.

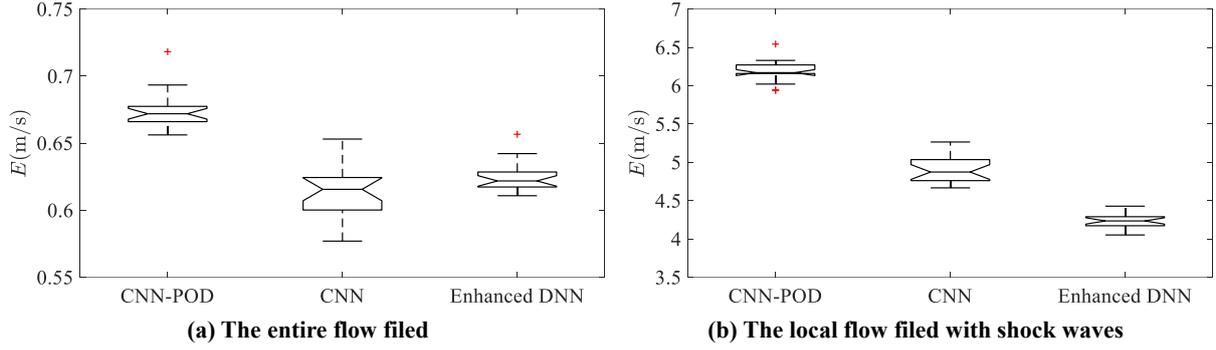

(a) The entire flow filed    (b) The local flow filed with shock waves

**Fig. 19 Box plots for the prediction error of streamwise velocity by different methods of 20 runs.**

Fig. 20 and Fig. 21 present the error contours and predicted crossflow velocity fields, respectively. The results in Fig. 21 show that the crossflow velocity field changes a smoother variation and lacks flow structures with significant gradients, such as the shock wave. Therefore, for all the prediction methods, the prediction performance of the crossflow velocity is superior to that of the streamwise velocity. All the prediction models demonstrate larger prediction error in the region around the shock wave. Among them, the CNN-POD model exhibits the best prediction accuracy, highlighting the effectiveness of the proposed method in improving prediction accuracy through the enhanced modeling strategy.

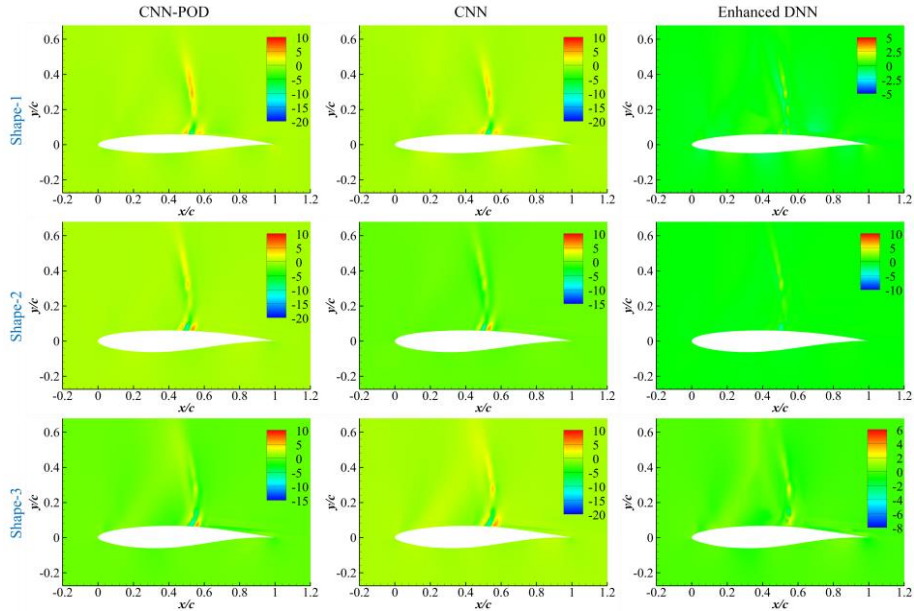

**Fig. 20 The prediction error of crossflow velocity by different methods for three airfoils, where the unit is m/s.**

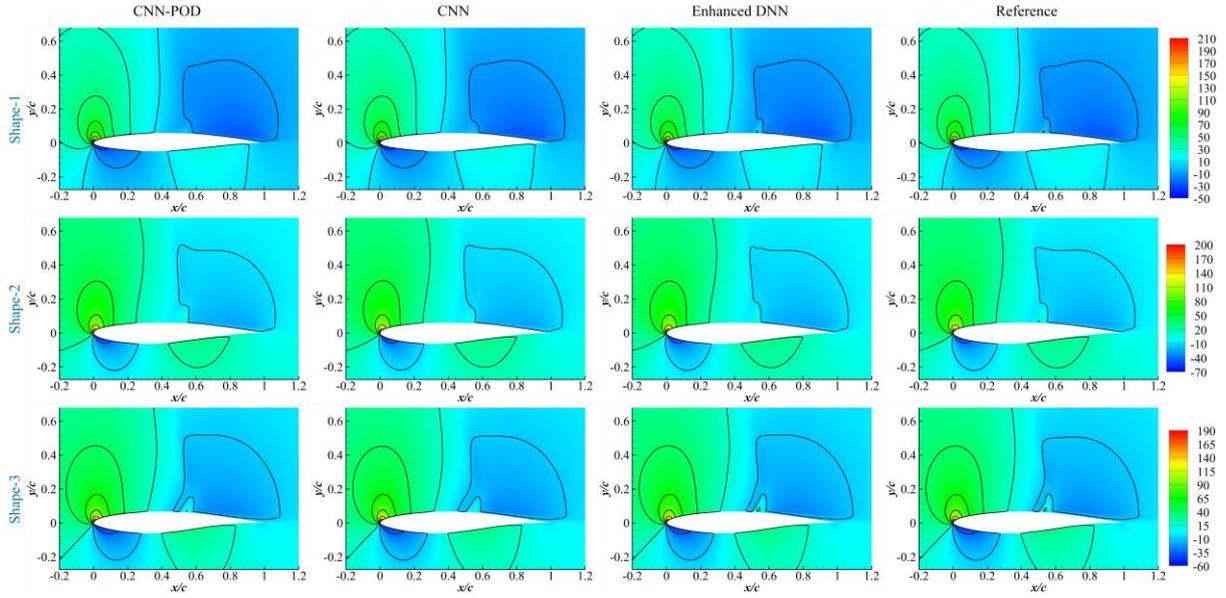

**Fig. 21 The forecasted results of crossflow velocity by different methods for three airfoils, where the unit is m/s.**

Table 4 presents the modeling parameters and performance indicators for crossflow velocity prediction using different methods. The enhanced DNN model exhibits the highest modeling accuracy in both the entire and local flow fields. Interestingly, when modeling the entire flow field, the CNN-POD model shows smaller prediction error compared to the CNN model, contrary to the results obtained for pressure and streamwise velocity prediction models. This difference can be attributed to the fact that crossflow velocity lacks strongly nonlinear flow structures, enabling accurate reduced-order modeling using POD. The enhanced DNN model continues to be effective in improving modeling accuracy of crossflow velocity field. In comparison with CNN-POD and CNN models, the enhanced DNN model achieves a reduction in prediction error of 10.53% and 26.09% for the entire flow field, and 46.20% and 46.20% for the local flow field, respectively. Additionally, the modeling time of the enhanced DNN model is 66.85% shorter than that of the CNN model, indicating the superiority of the proposed method in terms of both modeling accuracy and efficiency.

**Table 4 The modeling parameters and performance indicators for the prediction of crossflow velocity by different methods**

| Model | Input dimension | Output dimension | Parameters (M) | T (min) | $E$-entire (m/s) | $E$-local (m/s) |
|---|---|---|---|---|---|---|
| CNN-POD | 64×128 | 30 | 1.3 | 20 | 0.19 | 1.58 |
| CNN | 64×128 | 73904 | 328.5 | 178 | 0.23 | 1.58 |
| Enhanced DNN | 64×128 | 30+1682 | 1.3+7.0 | 20+39 | 0.17 | 0.85 |

All prediction models underwent 20 rounds of training, and the prediction results are depicted in Fig. 22. In terms of modeling of entire flow field, the CNN model exhibits a broader range of error distribution in comparison with the CNN-POD and enhanced DNN models, indicating a higher degree of training randomness in the CNN model. Conversely, for modeling the local flow field, all methods demonstrate good robustness, and the enhanced DNN model consistently achieves significantly smaller prediction error in each training round when compared to the CNN-POD and CNN models.

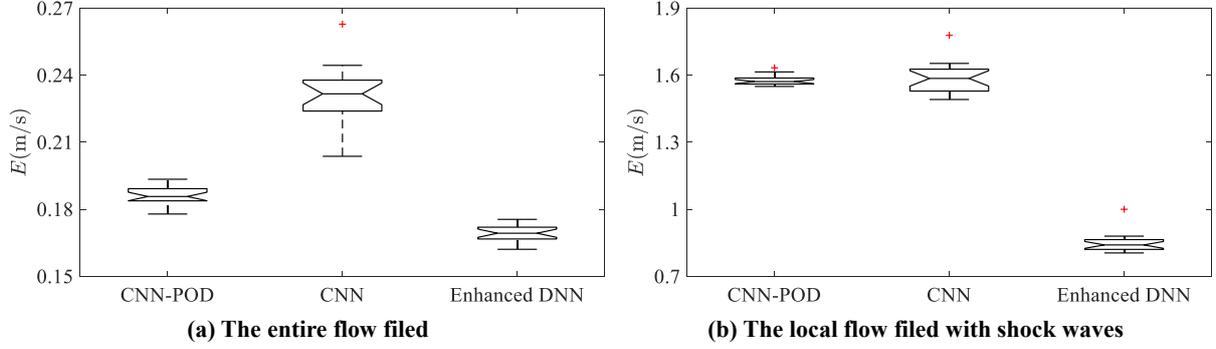

(a) The entire flow filed          (b) The local flow filed with shock waves

Fig. 22 Box plots for the prediction error of crossflow velocity by different methods of 20 runs.

In summary, the prediction results for pressure, streamwise velocity, and crossflow velocity fields consistently show that the proposed method outperforms the CNN-POD and CNN methods in terms of accuracy. Although the proposed method requires a longer training time and larger parameter size than the CNN-POD method, it is significantly more efficient than the CNN model in terms of both time and parameter size. The robustness of the propose method is comparable to the POD-CNN method and superior to the CNN method. Furthermore, the proposed method exhibits a high level of generalizability and can be effectively applied to different flow properties for enhanced modeling, regardless of whether the flow contains strongly nonlinear structures or not.

## 5. Conclusions

This work proposes an accuracy-enhanced flow prediction model by combining deep learning and reduced-order model. The proposed methos is validated by transonic flow prediction over airfoils. The following conclusions can be drawn from this study:

1. The proposed method can improve the modeling accuracy of the flow with strongly nonlinear flow structures. The enhanced DNN model outperforms the CNN-POD and CNN models by reducing the prediction error of the local flow, specifically around the shock wave, by 36% and 18% respectively in the pressure field. Moreover, the enhanced DNN method exhibits similar accuracy improvements when predicting the velocity fields.

2. The proposed enhanced DNN method shows better efficiency than CNN model. By employing the

enhanced DNN to capture the POD mode coefficients and local flow field, the modeling dimension is further reduced compared to the CNN model. The training time of the proposed model is approximately 1/3 of that in the CNN model. Additionally, the proposed model demonstrates similar robustness to the CNN-POD model and outperforms the CNN model in terms of the distribution of prediction error across repeated training.

3. For modeling flow fields that do not contain strongly nonlinear flow structures, such as the crossflow velocity field in this paper, the proposed method remains effective in improving the accuracy of flow field modeling. Moreover, the indirect modeling approach based on the CNN-POD model outperforms the direct modeling approach based on the CNN model in terms of both accuracy and efficiency.

**Declaration of competing interest**

The authors declare that they have no known competing financial interests or personal relationships that could have appeared to influence the work reported in this paper.

**Acknowledgment**

The research was supported by the National Natural Science Foundation of China (No. U2141254). The authors thankfully acknowledge this institution; the major advanced research project of Civil Aerospace from State Administration of Science Technology and Industry of China.

**Appendix A Validation of Grid independence**

The computational mesh for the reference airfoil RAE2822 is generated using the C-mesh, as shown in Fig. A.1(a). The total grid size is 73904 (496×149 layers). The first layer height of boundary layer grids is $5 \times 10^{-6}$ $c$. The partial computational mesh around the airfoil is shown in Fig. A.1(b). The upper surface and lower surfaces are discretized with 180 and 120 nodes, respectively. The mesh refinement of the upper surface is intended to obtain accurate flow field near the shock wave. Fig. A.2 compares the calculated pressure coefficient with the experimental data at $Ma = 0.734$, $\alpha = 2.79°$, and $\text{Re} = 6.5 \times 10^6$, it demonstrates good agreement between the calculated data and the experimental data.

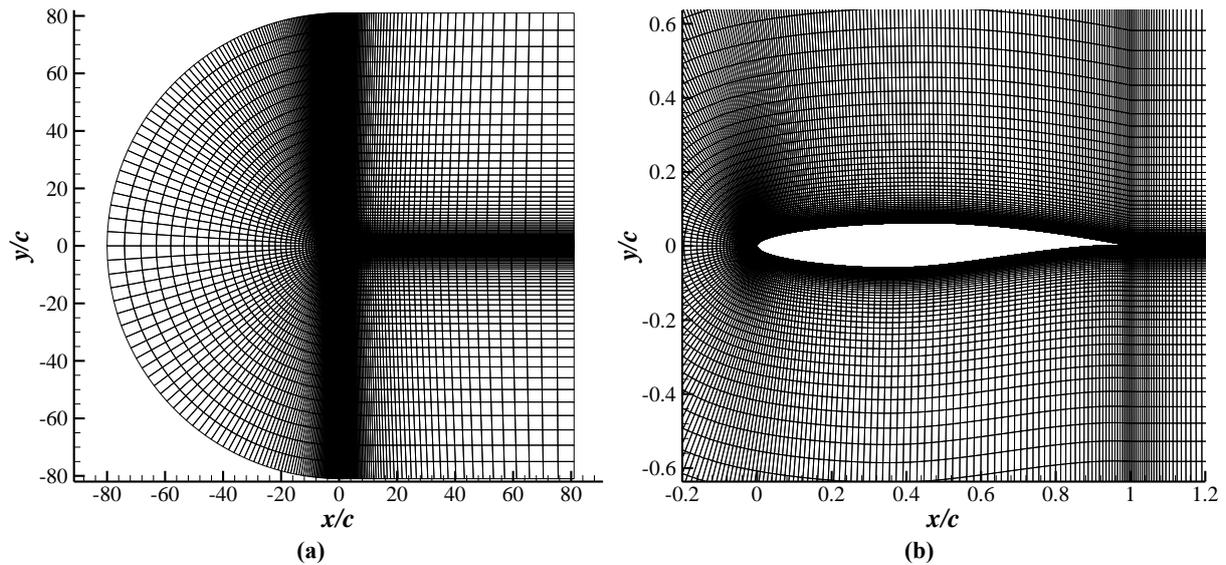

**Fig. A.1** The computational mesh of airfoil RAE2822. (a) The computational mesh; (b) The partial computational mesh.

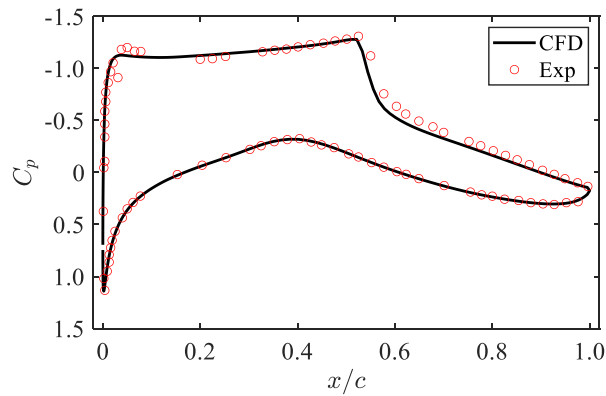

**Fig. A.2** The pressure coefficient distributions of airfoil RAE2822.